\documentclass[sigconf]{acmart}
\copyrightyear{2023}
\acmYear{2023}
\setcopyright{acmlicensed}
\acmConference[L@S '23] {Proceedings of the Tenth ACM Conference on Learning @ Scale}{July 20--22, 2023}{Copenhagen, Denmark.}
\acmBooktitle{Proceedings of the Tenth ACM Conference on Learning @ Scale (L@S '23), July 20--22, 2023, Copenhagen, Denmark}
\acmPrice{15.00}
\acmISBN{979-8-4007-0025-5/23/07}
\acmDOI{10.1145/3573051.3593377}

\begin{document}

\title[Sharing Educational Data Science Research With School Districts]{All A-board: Sharing Educational Data Science Research With School Districts}

\author{Nabeel Gillani}
\authornote{Denotes contributions of equal importance.}
\affiliation{%
  \institution{Northeastern University}
  \city{Boston, MA}
  \country{USA}}
    \email{n.gillani@northeastern.edu}

\author{Doug Beeferman}
\authornotemark[1]
\affiliation{%
  \institution{Massachusetts Institute of Technology}
  \city{Cambridge, MA}
  \country{USA}}
\email{dougb5@mit.edu}

\author{Cassandra Overney}
\affiliation{%
  \institution{Massachusetts Institute of Technology}
  \city{Cambridge, MA}
  \country{USA}}
\email{coverney@mit.edu}

\author{Christine Vega-Pourheydarian}
\affiliation{%
  \institution{Wellesley College}
  \city{Wellesley, MA}
  \country{USA}}
\email{cpourhey@wellesley.edu}

\author{Deb Roy}
\affiliation{%
  \institution{Massachusetts Institute of Technology}
  \city{Cambridge, MA}
  \country{USA}}
\email{dkroy@media.mit.edu}

% \renewcommand{\shortauthors}{Gillani and Beeferman et al.}
\renewcommand{\shortauthors}{Nabeel Gillani, Doug Beeferman, Cassandra Overney, Christine Vega-Pourheydarian, \& Deb Roy}

\begin{abstract}
  Educational data scientists often conduct research with the hopes of translating findings into lasting change through policy, civil society, or other channels.  However, the bridge from research to practice can be fraught with sociopolitical frictions that impede, or altogether block, such translations---especially when they are contentious or otherwise difficult to achieve.  Focusing on one entrenched educational equity issue in US public schools---racial and ethnic segregation---we conduct randomized email outreach experiments and surveys to explore how local school districts respond to algorithmically-generated school catchment areas (``attendance boundaries'') designed to foster more diverse and integrated schools.  Cold email outreach to approximately 4,320 elected school board members across over 800 school districts informing them of potential boundary changes reveals a large average open rate of nearly 40\%, but a relatively small click-through rate of 2.5\% to an interactive dashboard depicting such changes.  Board members, however, appear responsive to different messaging techniques---particularly those that dovetail issues of racial and ethnic diversity with other top-of-mind issues (like school capacity planning).  On the other hand, media coverage of the research drives more dashboard engagement, especially in more segregated districts.  A small but rich set of survey responses from school board and community members across several districts identify data and operational bottlenecks to implementing boundary changes to foster more diverse schools, but also share affirmative comments on the potential viability of such changes. Together, our findings may support educational data scientists in more effectively disseminating research that aims to bridge educational inequalities through systems-level change.
\end{abstract}





\maketitle

\section{Introduction}

There is perhaps no place in America where failures of pluralism and democracy are manifesting more clearly than in public schools.  In recent years, US school districts have become battlegrounds for political debates about in-person schooling~\cite{sawchuk2021boards}, mask mandates~\cite{harris2022covid}, and critical race theory, along with discussions about gender and sexuality~\cite{natanson2021loudoun}.  Despite this climate, public schools have the capacity to help cultivate common ground, more evenly distribute social capital, spark exposure to and appreciation for divergent viewpoints, and foster other aspects of a healthy, pluralistic society by bringing together students and families from different backgrounds.  

Yet across the US, school segregation along racial, cultural, and socioeconomic lines remains rampant~\cite{gao2022segregation} and continues to adversely affect the academic performance and attainment of lower socioeconomic status (SES) students and students of color~\cite{reardon2018testgaps,owens2018income,reardon2014seg,mayer2002income}.  One reason for this is that segregation may exacerbate inequities in how resources like experienced teachers and advanced course offerings are distributed across schools~\cite{edbuild2020lines,frankenberg2003civil,orfield2005civil}.  Segregated schools may also impede access to social networks, which may in turn limit access to the kinds of ``bridging social capital'' that can help them access new resources, jobs, and other quality-of-life-enhancing opportunities~\cite{bellInventor,smallUnanticipatedGains,burton2015inequality,chetty2022socialcapitalII}.  There is evidence that integration can reduce inequalities in academic outcomes~\cite{yeung2022france,quick2016cps, johnson2011desegregation}, especially when academic environments are re-designed to be inclusive of an increasingly diverse student body~\cite{burton2015inequality,bridges2016eden}.  

Despite such evidence, actual socioeconomic integration remains elusive across many districts, largely because of racialized schooling preferences among predominantly White and affluent families~\cite{billingham2016parents,hailey2021parents,hall2017migration} and residential/school selection based on factors related to income~\cite{reardonIncomeInequality,hasanDigitizationDivergence}.  The correlation between place of residence and place of schooling is reinforced by ``school attendance boundaries'', or school catchment areas that local education agencies (``school districts'') draw to determine which neighborhoods are assigned to (``zoned for'') which schools.  Despite the growing popularity of school choice policies, neighborhood-based assignment still accounts for the vast majority of student assignment decisions across the US~\cite{nces2021choicefacts}.  Furthermore, choice policies can also exacerbate patterns of racial and ethnic segregation~\cite{monarrez2022charters,whitehurt2017segregation,candipan2019neighborhoods}---despite how they, at least in theory, place of residence and schools.  Many choice systems also factor in geography in some form or fashion---for example, through the priorities they assign students~\cite{monarrez2021urban} or how they create zones that constrain choice sets for families~\cite{campos2022zones}.  Boundaries, therefore, matter, and play an important role in determining which children go to which schools and what the quality of their ensuing educational experiences are.

Educational data scientists---an emerging class of researchers and practitioners wielding computational methods to explore outstanding issues in education~\cite{mcfarland2021eds}---have expended great effort exploring how such methods can improve online learning~\cite{kizilcec2013mooc,gillani2014mooc}, personalized and adaptive tutoring~\cite{ritter2007its,piech2015rnn}, and other learner and teacher-centric practices.  There appears to be fewer studies, however, investigating how such methods might help address challenges that are ``upstream'' to these: namely, social, organizational and contextual barriers that might impede access to quality-of-life enhancing educational opportunities.  Attendance boundaries that exacerbate segregation often act as such barriers.  Yet they are malleable: school districts could, in theory, redraw attendance boundaries in ways that seek to minimize racial and ethnic segregation across schools.  Doing so, however, is a politically challenging task, and often runs the risk of increasing travel times or producing other unpopular outcomes that face opposition from parents and other stakeholders~\cite{delmont2016busing,mcmillan2018boundaries}.  It is also technically challenging: manual map redrawing can be highly time-consuming given the large space of possible configurations, and from the political redistricting literature, it is clear that automated methods are challenging to develop and scale in their own right~\cite{gurnee2021fairmandering,becker2020redistricting}.  There is an opportunity, then, for educational data scientists to explore how computational methods might help education policymakers---a stakeholder group not usually directly targeted as an audience for pedagogy-focused research---foster more diverse and integrated schools.  This is, arguably, especially relevant for educational data scientists interested in ``Learning at Scale'', since the school assignment policies that districts design and implement continue to shape the learning experiences of millions of children across the US.  

In this vein, a recent study developed computational redistricting algorithms to explore how much school districts across the US might reduce racial and ethnic segregation in elementary schools through boundary changes~\cite{gillani2023redrawing}.  The results, surprisingly, revealed the possibility of boundary changes that might reduce segregation while also slightly reducing travel times.  These findings motivate this particular study: a follow-on effort to explore to what extent US school districts might engage with the findings from this research and take them into consideration as they conduct their own boundary planning efforts.  To investigate this, we adopt a multi-pronged outreach strategy, combining cold email outreach to 4,320 elected school board members across 803 districts with media coverage in order to invite leaders and members of school districts to explore an online dashboard depicting algorithmically-generated, diversity-promoting elementary school attendance boundaries.  Our focus on outreach seeks to address another gap, not just in educational data science research, but academic research more broadly: better understanding the conditions under which academic findings might be translated into practice.  The emerging field of ``implementation science'' is increasingly exploring this question~\cite{bauer2015implementation}, and recent studies in the education domain have used school district outreach as a channel~\cite{delevoye2021covid,nakajima2021evidence} for investigating how district leaders interpret and put policy-relevant research to use.  We aim to build on these bodies of work by asking the following questions:

\begin{enumerate}
    \item Which messaging and outreach strategies are more or less effective in prompting school district leaders and community members in exploring potential boundary changes that could help foster more racially/ethnically integrated schools?
    \item How do school district leaders and community members perceive these boundary changes---namely, what potential merits and pitfalls do they see in both how these boundaries were produced and their potential viability in practice?
\end{enumerate}

We find that email open and click rates are highest when campaign content is framed in terms of existing district objectives---like school capacity planning and family satisfaction---which were suggested by district leaders through our informal conversations with them.  Furthermore, media coverage appears to be a much more effective channel than email outreach in engaging audiences around potential boundary changes in their school districts---especially districts that exhibit higher levels of White/non-White segregation.  Through an optional survey available on the data dashboard, we obtain a more nuanced view of how communities perceive such hypothetical, algorithmically-informed attendance policy changes: many participants highlight aspects of their unique contexts that would need to be considered in order for the research findings to be adapted and eventually implemented in practice, like rapid shifts in population due to immigration; school capacity constraints; and other factors.  Nevertheless, most respondents indicate more interest in the topic of attendance boundary changes to foster diverse schools than they had prior to learning about the study, and several express an interest in learning more about how to implement some of the proposed changes in practice.  While our study focuses on the issue of changing attendance boundaries to foster more racial and ethnic integration, we believe our findings may be relevant more broadly to educational data scientists who wish to explore how findings from research can be effectively communicated with key education stakeholders, and eventually translated into effective socio-technical artifacts and practices in learning settings.

\section{Related work}

Our work builds upon work in school assignment planning for measuring and promoting more diverse schools.  It also builds upon recent efforts exploring how to engage school district leaders and other policymakers to help them use findings from academic research in their work. Most broadly, it is motivated by and rooted in the emerging discipline of ``implementation science''~\cite{bauer2015implementation}, which seeks to more precisely identify the contexts and conditions under which research findings can be scaled and generalized to populations beyond those included in the research. We briefly review these areas below.

\subsection{School assignment for diversity}

Across school districts, the lines \textit{between} districts have been shown to perpetuate racial and ethnic segregation across schools even more than those \textit{within} them (i.e., attendance boundaries)~\cite{fiel2013boundaries}. Unfortunately, changing school district boundaries require collaboration across multiple districts, and/or state-level intervention---both of which are subject to many of the political frictions and challenges that accompany policy changes requiring coordination across multiple governing bodies.  Changing attendance boundaries, on the other hand, generally falls under the purview of school district leaders.  Yet within-district boundary changes continue to be highly contentious: parents may worry about impacts on values~\cite{kane2005housing, bridges2016eden, black1999housing}, commute times~\cite{frankenberg2011polls, delmont2016busing}, community cohesion~\cite{bridges2016eden, baltimore2019}, and other factors.  These concerns can block boundary changes~\cite{castro2022richmond} and/or spark families to move to other schools and districts~\cite{reber2005flight,macartney2018boards}.  

The present study does not explore how to address these parent/ community-level factors, which are critical topics for future research on this issue.  Instead, it explores to what extent school district leaders are willing even to engage with research on this contentious topic that suggests there might be potentially palatable solutions within reach---and what their reactions to such findings are.  Given the positions of power that board members occupy, we believe this is important to investigate, as identifying effective ways of discussing the topic of boundary changes for diversity with district leaders may illuminate key leverage points for supporting systemic policy change.

\subsection{Engaging education policymakers in research findings}
While still a relatively nascent area of inquiry, recent studies have started to explore how to design and execute effective communications with school districts to foster more engagement with, and policy decision-making based on, evidence-based research.  Two recent studies in particular come to mind.  The first conducted randomized outreach to over 2,000 education professionals working across both state and local education agencies to assess both the preferences these policymakers have for research evidence, and the extent to which they update their beliefs about the effectiveness of various education policies as they are presented with new evidence~\cite{nakajima2021evidence}.  The study found that policymakers do indeed value research evidence, particularly when they contain findings from multiple sites, large samples, and contexts similar to their own---though they do not appear to prefer experimental studies over observational ones.  The study also found that policymakers are more likely to update their beliefs about the effectiveness of education policies when research evidence is presented with clear, accessible descriptions of its research design.  Together, these results offer valuable insights into how clarity of communication and local contextualization of findings might help increase policymakers' engagement with education research.

A second study, conducted during the COVID-19 pandemic, explored the potential agenda-setting role of cold email outreach to school district board members and staff~\cite{delevoye2021covid}.  The authors randomly assigned a subset of nearly 600 school districts leaders to receive an email containing a link to a memo with research on emerging best practices around transportation in education settings during the pandemic.  They then analyzed video recordings and other outputs from school board meetings to analyze whether those district leaders who received emails were more likely to discuss transportation-related issues.  While the study did not report a positive treatment effect, it did find that some of the words from the memo made it into some of the school boards' discourse.  It is possible that with a larger sample of districts, there may have been more of an agenda-setting impact of such research-based outreach.  

Our study builds on these efforts by exploring the types of cold email outreach that are more or less effective at garnering engagement from school district leaders vis-a-vis a particularly contentious and systemic issue: how school attendance boundaries are drawn, and the impact these boundaries have on issues of school segregation and diversity in their districts.

\subsection{Implementation science}
Yet another emerging area of inquiry is implementation science, or “the scientific study of methods to promote the systematic uptake of research findings and other evidence-based practices into routine practice ...”~\cite{bauer2015implementation}.  For example, development economists are increasingly exploring implementation science through the lens of ``scaling'', or investigating how promising interventions conducted in relatively sand-boxed settings might be effectively scaled across different contexts and environments~\cite{mobarak2022devecon}.  Scaling successfully, of course, requires anticipating the complexities that may come with scale---which, crucially, include potentially unanticipated or indirect consequences~\cite{mobarak2022devecon}.  

Health researchers are also interested in implementation science; indeed, the health domain has been a driver of this new discipline, perhaps because evidence-based practices in healthcare take, on average, 17 years to be implemented into routine general practice~\cite{bauer2015implementation}.  One interesting emerging trend from this domain is the design of ``hybrid research designs'', or research projects that seek to both further understanding about evidence-based practices (e.g., how well some practice improves a health outcome---generally, the type of finding one might typically expect to uncover by running a randomized control trial), as well as a documentation/investigation of which practices and policies are most effective in enabling such interventions to have this intended impact~\cite{bauer2015implementation}.  Such hybrid designs offer the hope of both advancing basic scientific inquiry as well as theories and knowledge about how the outputs of such inquiry can actually serve people in different real-world contexts.

Many of these ideas have started to take hold in education research~\cite{moir2018edimplem}, as partly demonstrated by the district outreach studies in the prior subsection.  They are also beginning to spark new field-building and funding priorities in education research~\cite{grant2022evidence}.  For an issue as complex as school assignment planning, particularly when intersected with a topic as controversial as racial and ethnic segregation in schools, modeling hypothetical boundary changes is just one small part---and perhaps the easiest part---of investigating how to foster more diverse and integrated schools.  Deeply understanding the sociopolitical contexts of these districts, how to capture the attention and interest of district leaders and community members, and eliciting the conditions under which alternative boundaries may or may not be viable are all critical pieces of translating such research into practice.  On their own, these activities do not actually identify conditions for successful translations in the field, but rather, help identify potential starting points for iterating on existing research to eventually help make it more useful in practice.  Our study, then, may offer helpful insights to educational data scientists, public policy researchers, and other practitioners interested in learning more about how to begin translating potentially controversial, equity-promoting research findings into practice in complex settings like school districts.

\section{Methods}

\subsection{Modeling and data dashboard}
We begin with the datasets and rezoning algorithms used in~\cite{gillani2023redrawing}.  This preliminary study focused on elementary school segregation, combining 2019/2020 student enrollment data by race and ethnicity across US schools with 2020 Census demographics and 2021/2022 school attendance boundaries to estimate counts of students living in each Census block and attending each zoned elementary school (see~\cite{gillani2023redrawing} for a more thorough discussion of the datasets and some of the challenges involved in making these estimations).  It then developed and presented a redistricting algorithm, building on top of open source tools from the operations research community~\cite{cpsat,cpsat2020youtube}, that produced hypothetical attendance boundary redrawings to help foster more racially and ethnically integrated schools.  The redrawing problem was framed as an assignment problem, namely, trying to find an alternative assignment of Census blocks to schools that would minimize White/non-White segregation (as defined by the dissimilarity index~\cite{massey1988segregation}), subject to constraints around the maximum amount any family's travel time might increase under the new rezoning, as well as a cap on how much school capacities might increase.  Such redistricting problems are NP-hard and notoriously difficult to solve given the large space of possible assignments; nevertheless, recent advances in the operations research community, e.g. through innovations in methods like constraint programming~\cite{pascal1989cp, cpsat2020youtube}, have created powerful off-the-shelf solvers that can often produce high-quality solutions to such problems. 

\cite{gillani2023redrawing} simulated changes across 98 large school districts and reported a median relative decrease in segregation of 12\%, which would require nearly 20\% of students to switch schools.  Perhaps most surprisingly, the study showed that such reductions may be achieved with slight average \textit{decreases} in travel times---suggesting there may be modest changes that produce modest gains in integration without requiring large travel-related sacrifices by families.

Using the datasets and algorithms from this study, we began by re-running the authors' redistricting algorithms for a larger set of over 4,000 school districts---namely, those school districts that have at least two elementary schools (and hence, different neighborhoods within the district zoned to different schools).  In total, we produced 4 different configurations across each district by varying different parameter values for the aforementioned travel time and school size constraints.  Simulations were run on a university computing cluster with a cutoff time of 5.5 hours each.  We then produced a dashboard similar to the one in~\cite{gillani2023redrawing} to enable district leaders and community members to explore hypothetical rezonings.  The dashboard was built using the Python-based Streamlit\footnote{\url{https://streamlit.io/}.}, a lightweight framework for building and deploying data dashboards.  Using Streamlit limited some of our data tracking, logging, and customization possibilities; on the other hand, it enabled us to rapidly prototype and deploy a dashboard without requiring full-stack design and engineering.

\begin{figure*}
\centering
\includegraphics[width=0.8\textwidth]{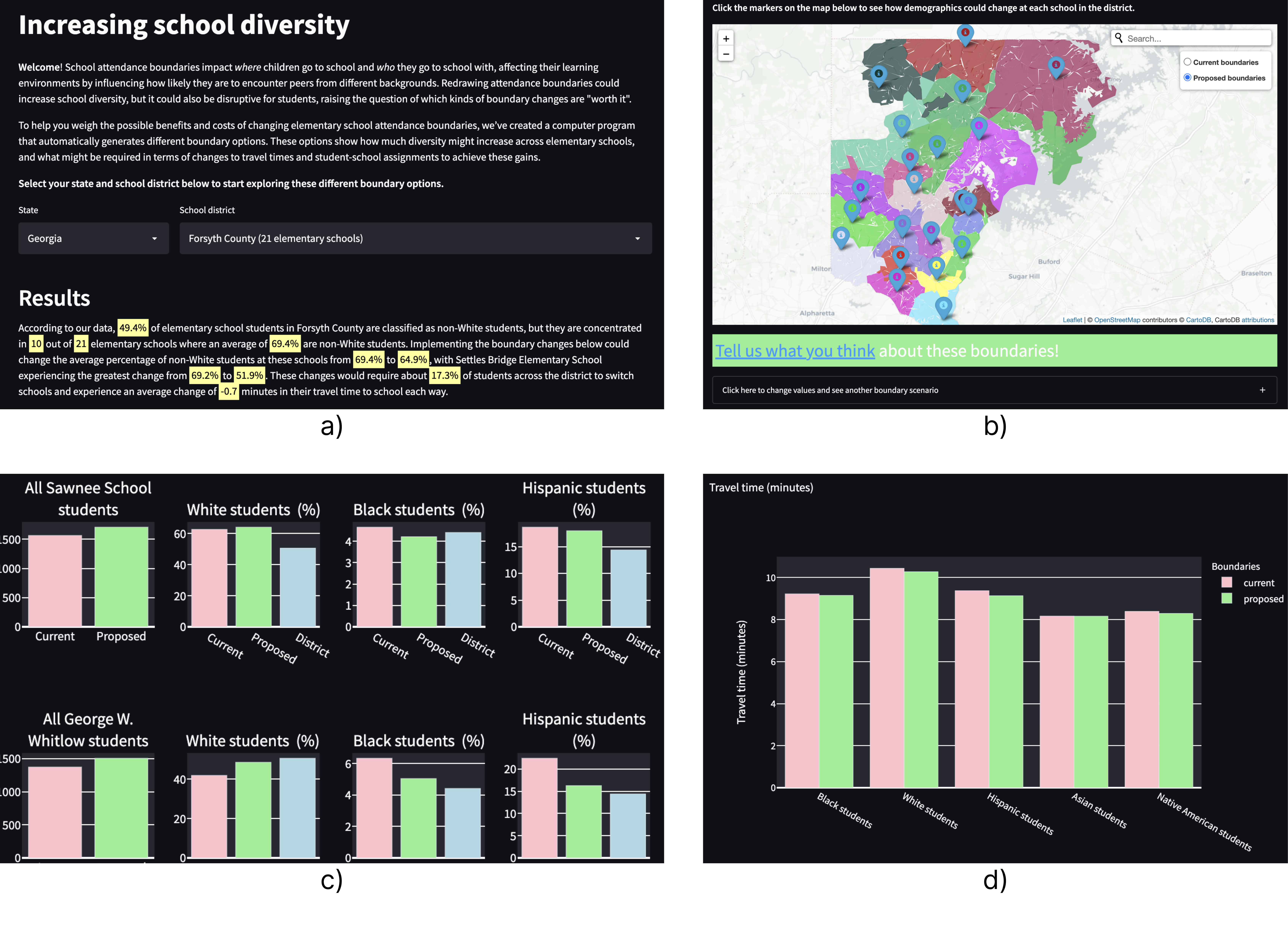}
\caption{Key screenshots of rezoning dashboard, which can be found at \url{https://www.schooldiversity.org/}: a) shows the view ``above the fold'' when users land on the page and includes a description of why attendance boundaries matter and what the objective of the dashboard is.  It also provides dropdown boxes for users to explore hypothetical boundary changes for different school districts, and includes a verbal description of what the depicted changes for any given district would mean in terms of changes in diversity across its constituent elementary schools.  b) shows a map view, where users can toggle between boundaries proposed by the redistricting algorithm and current ``status-quo'' boundaries.  Further down the page and hidden by default is c), which allows users to drill into specific schools to understand how their demographics would change under the depicted rezoning---including cases where rezonings move racial/ethnic proportions at some schools \textit{away} from district-levels instead of towards them, perhaps to achieve more balance at other schools.  Finally, further down is d), which shows how travel times would change (on average) for different racial/ethnic groups under the rezoning.}
\label{fig:1}
\end{figure*}

Figure~\ref{fig:1} shows screenshots of the dashboard, which served as the main exploration platform for this study. 
 To make the description of the project's aims and key findings as accessible as possible on the dashboard, we informally sought feedback from a past school district superintendent involved in boundary planning for integration; a current (at the time) school assignment and planning director; and a data journalist who frequently writes about issues relating to educational equity.  We conducted Zoom calls with these individuals, inviting them to explore the dashboard and share their observations and questions (similar to a think-aloud protocol~\cite{ericsson1984think}).  Their feedback on clarity of wording (how we describe changes in school-level demographics---and avoid esoteric segregation formulas in order to make research findings more accessible~\cite{nakajima2021evidence}), framing of diversity (making sure to not lose the focus on within-classroom learning experiences) and data visualizations was invaluable and led us to make several changes that we believe made the dashboard more accessible to a general audience.

\subsection{Email campaign}
Since school board members across many districts are elected officials, their names and contact information are available through public ballot databases.  We collaborate with the XQ Institute\footnote{\url{https://xqsuperschool.org/}.} to obtain this information for over 115k elected officials across 13,233 school districts---which constitute nearly all regular school districts across the US\footnote{\url{https://nces.ed.gov/programs/digest/d20/tables/dt20_214.10.asp}.}.  Next, via the elementary school attendance boundaries and school enrollment data used in~\cite{gillani2023redrawing}, we identify the 1,000 school districts with the largest elementary school enrollment \textit{that are also listed as having closed enrollment (i.e., entirely boundary-assigned) elementary schools} as our main sample for outreach.  Merging the district board member data onto this dataset yields approximately 6,664 board members, 4,831 of which we have a valid email address for.  We then design and send emails using the email marketing platform Mailchimp.  Emails contain links unique to each recipient, and clicking the link takes the recipient to a preliminary page on the dashboard that informs them of the research study.  The recipient is then invited to click through to the main dashboard, where they are shown the dashboard depicted in Figure~\ref{fig:1} with data specific to their district pre-selected.

As described in the Introduction by Research Question 1, we are interested in better understanding which messaging techniques are more or less effective in prompting school district leaders (board members) to explore and share feedback on potential attendance boundary changes for school diversity.  One way to do this is to randomly assign board members to receive different emails---e.g., emails with different subject lines---and then observe which ones generate more engagement and eventual traffic to the data dashboard (we use ``email'' and ``subject line'' below interchangeably, since email previews and bodies were generally similar across runs and only adapted to reflect their corresponding subject lines).  

Randomly assigning board members within the same district to different email conditions, however, runs the risk of treatment ``spillover''---for example, if one board member forwards an email to one of their colleagues---which would limit our ability to identify the causal effect of any particular email on engagement~\cite{gerber2012fieldexpar}.  To try and mitigate against this, we design a series of cluster-randomized control trials, where we randomly assign email conditions at the cluster (school district) level.  There is still the possibility that school board members across districts who are assigned to different conditions forward received emails to one another, but we expect this to be small, given their specific contexts (boundaries) are likely to be different.  To account for the fact that board members might respond differently depending on the extent to which demographic imbalances are actually an issue in their particular districts, we stratify districts into quintiles according to their White/non-White dissimilarity indices, and randomly assign those within each quintile to each possible email condition.  This increases the likelihood that districts receiving any of the email conditions are well-balanced across the levels of segregation manifesting across their schools.

We conduct these cluster randomized control trials in eight successive waves.  At the start of each wave, we begin with a set of hypotheses for which types of subject lines would be most effective in driving open and click-through rates.  After each wave, we analyze engagement data (opens and click-throughs to the dashboard) for each email.  The next wave then contains emails informed by well-performing ones in prior waves, plus those evaluating new hypotheses.  We also generate new hypotheses---based on observations and informal interviews conducted with school board members who engaged in previous waves---to inform the development of new subject lines.  In the results section, we discuss some of these interviews and the insights that emerged from them.  Table~\ref{tab:1} contains some of our hypotheses and corresponding subject lines, with a full list deferred to Appendix~\ref{app:subjects}.  In total, we explored 14 subject lines across 17 campaigns, which were sent through 8 iterative waves.  The median number of recipients across campaigns was 205.  Our final email included a subject line that included aspects of well-performing emails in prior waves; we sent this to 1,383 participants that remained in our total sample after running all previous subject line experiments.

We note that due to the exploratory and iterative nature of this study, we did not formally pre-register it in an Open Science repository.  We acknowledge this as a limitation of this work, as even exploratory studies and sequential experimentation can be accommodated in existing pre-registration frameworks~\cite{nosek2018prereg}, and intend to rectify this in future studies of this type.

\begin{table}[ht]
    \centering
    \begin{tabular}{p{0.5\linewidth} | p{0.5\linewidth}}
      \textbf{Email subject line} & \textbf{Hypothesis} \\ \hline
      \$20 for your views on diversifying attendance boundaries & A small incentive that board members can allocate to a teacher of their choice may increase engagement \\
      Data science to improve learning experiences & Using a popular term (``data science'') and focusing on learning instead of diversity may increase engagement \\
      More diversity with shorter commutes? & Stating the main finding of~\cite{gillani2023redrawing} up front may increase engagement \\
      Data science to decide which schools to open or close & Framing in terms of a problem school districts often face that triggers boundary planning may increase engagement \\
    \end{tabular}
    \caption{Example email subject lines and corresponding hypotheses motivating their inclusion across different email waves.}
    \label{tab:1}
\end{table}

We use logistic regression with cluster robust standard errors to analyze the causal effect of each subject line on open and click-through rates.  The use of cluster robust standard errors is typical in such types of experimental designs and generally meant to account for correlations between participants (board members) in the same cluster (district) that may impact results~\cite{cameron2015cluster}.  We control for several variables that may also affect the propensity for board members to engage with such outreach.  The regression specification is as follows:

\[
y^o_i = \beta_0 + \beta_1 \cdot S_i +  \beta_2 \cdot R_i + \Sigma^{8}_{j=3}~\beta_{j} \cdot D^j_{i} + \Sigma^{10}_{j=9}~\beta{j} \cdot C^j_{i} + \epsilon_i
\]

Where $i$ represents each recipient; $y^o_i$ represents the binary result for each outcome $o \in \{opened, clicked\}$; $S_i$ represents the subject line (campaign) randomly assigned to $i$; $R_i$ is an estimate of $i$'s race/ethnicity using the rethnicity package in R~\cite{xie2022rethnicity}; $D_i^{3:8}$ represent a set of district-level characteristics for the district $i$ belongs to; $C_i^{9:10}$ represent campaign-specific variables---namely, the day of the week and month (May or June) that the campaign was sent during, given potential impacts of week and school year-level seasonalities on engagement; $\beta_0$ is the intercept; and $\epsilon_i$ is the error term.  In particular, the variables $D_i^{3:8}$ represent: an indicator describing whether the school board is whiter than the district, suggesting a lack of racial representativeness in the district leadership; the district's White/non-White dissimilarity across elementary schools; the percentage of students in the district who are White; the district's urbanicity (urban, suburban, small city, rural); the number of elementary students enrolled across its schools; and its total number of elementary schools.  Once the model is fit, the coefficient $\beta_1$ indicates the causal effect of the particular email subject line on open and click-through rates, and the other coefficients indicate associations between various recipient, district, and campaign-level variables and these outcomes measures.   

\subsection{Survey}
We included an optional survey linked off of the data dashboard to invite leaders and community members to share their feedback on the depicted rezonings.  The main objective of the survey was to help answer Research Question 2, namely: ``How do school district leaders and community members perceive these boundary changes---namely, what potential merits and pitfalls do they see in both how these boundaries were produced and their potential viability in practice?''.  Survey questions assessed respondents' interest in the topic of boundary planning; their views of the algorithms that produced the hypothetical boundary scenarios; what they found interesting about the depicted scenarios; what they found concerning about the depicted scenarios; and space to leave any other questions or comments.  A full list of questions can be found in Appendix~\ref{app:survey}.

\section{Results}

\subsection{Email campaign}
We begin with a few descriptive statistics about the recipients and email campaigns before exploring results from our regressions.  Out of 4,831 board members included in our sample, emails were delivered to nearly 90\% (4,320); approximately 10\% of emails bounced (due to email address inaccuracies, being classified as spam, or other reasons) and were not delivered to the intended inbox.  Two-thirds of school districts in our sample had school boards with a Whiter racial composition than their elementary school student body. 
 We caveat this finding given potential biases and inaccuracies in race classification~\cite{lockhart2023race} (the rethnicity package has an overall f1 score of 0.78~\cite{xie2022rethnicity}).  To the extent it is correct, however, it reflects a broader trend of school boards failing to reflect the racial and ethnic compositions of ever-diversifying student and family compositions~\cite{samuels2020diversity}. 

In terms of email engagement, 39\% of recipients (1,686) opened the email they received, but only 2.5\% (106) clicked through to the dashboard.  Analyzing log data on the dashboard revealed that the vast majority of these click-throughs did not make it past the IRB-required text included on the landing page that describes the research study and data (see Appendix~\ref{app:irb} for this text).  Finally, just under 1.5\% of recipients (64) unsubscribed from receiving future emails.  Benchmarking open, click-through, and unsubscribe rates is generally difficult as they may vary across industries, purpose/type of campaign, and many other variables.  Nevertheless, using average estimates produced by Mailchimp---the platform we used for our email campaigns---an open rate of 39\% vastly exceeds their reported average of 23\%~\footnote{\url{https://mailchimp.com/resources/email-marketing-benchmarks/}.}.  Our observed 2.5\% click-through rate, however, is slightly below their reported average of 2.9\%.  Finally, while our 1.5\% unsubscribe rate is small in an absolute sense, it is an order of magnitude higher than their average of 0.25\%.  We caveat these Mailchimp benchmark numbers as only rough guideposts, given they represent averages over a wide range of campaigns and audiences. 
 Interestingly, our observed open rate is nearly identical to the 40\% open rate observed in~\cite{delevoye2021covid}, which involved outreach to school board members during COVID, though their click-through rate of 7\% is much larger than ours.  This may be due to the fact that their call to action involved reading a memo sharing information about school transportation during an evolving pandemic schooling context, while ours contained an invitation to explore findings pertaining to a much more politically contentious, and temporally less-pressing, issue.

\begin{figure*}
\centering
\includegraphics[width=.95\textwidth]{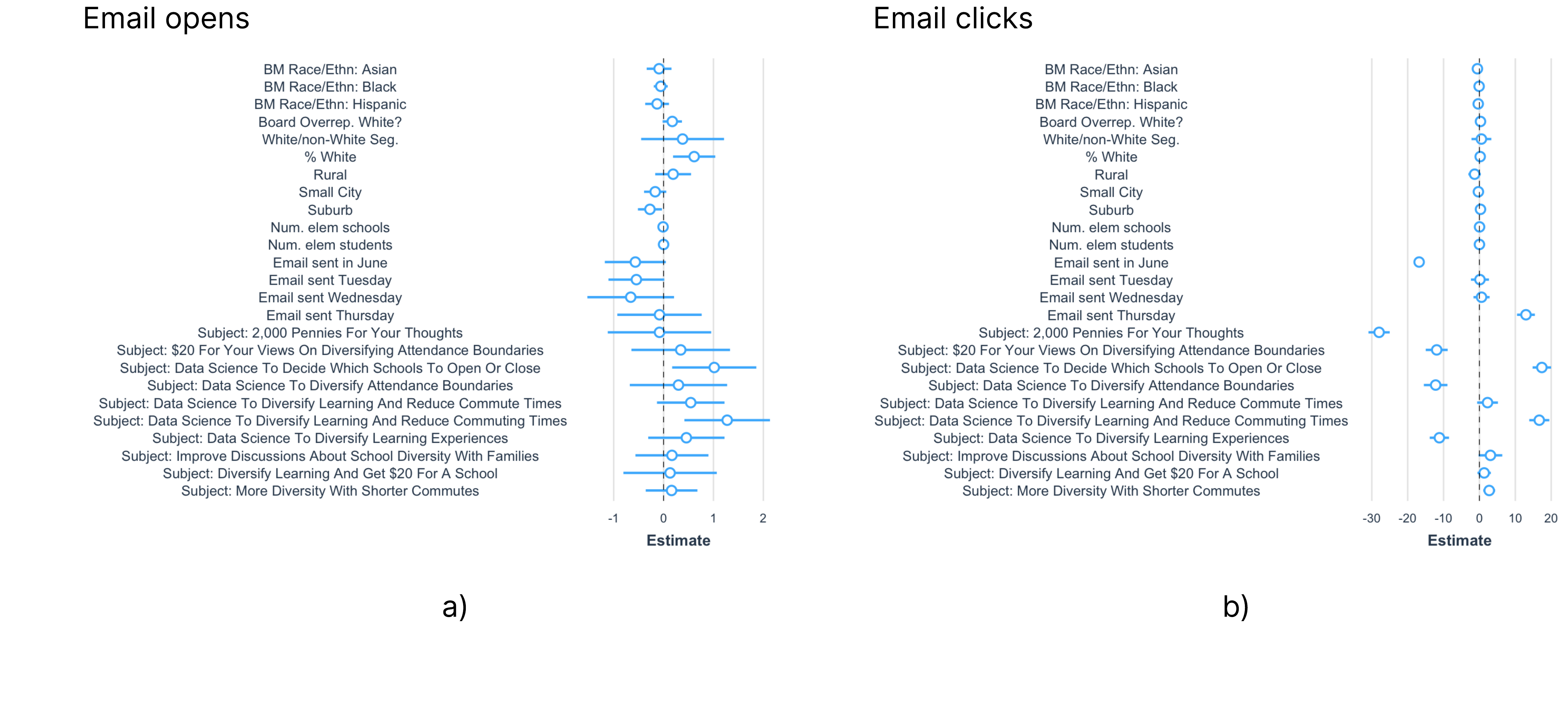}
\caption{Associations between different regression variables and likelihood of board members opening and clicking emails, respectively.  The ``Subject: ...'' variables represent different email campaigns and the depicted values can be interpreted as the magnitude of the causal effect of each campaign on open and click outcomes, since participants were randomly assigned to campaigns.  Circles represent average associations (i.e., regression coefficient values, which indicate log odds), and lines represent 95\% confidence intervals.  Intervals that do not intersect zero indicate a statistically significant association between that variable and its depicted outcome at an $\alpha$ level of 0.05.  The results for categorical variables indicate the magnitude of associations relative to a particular reference category: for BM Race, this is ``White''; for Urbanicity, it is ``urban''; for day of week email was sent, it is ``Monday''; for month email was sent, it is ``May''; and for the campaigns, it is the control subject line of ``Diversify Learning''.}
\label{fig:2}
\end{figure*}

Next, we analyze the results of our regressions to investigate Research Question 1 (``Which messaging and outreach strategies are more or less effective?'').  First, we analyze the causal effect of different email subject lines on the likelihood that recipients open our outreach emails, controlling for the additional individual, district, and campaign-level variables described in the Methods section.  Figure~\ref{fig:2}(a) shows that campaigns with the subject line ``Data Science To Decide Which Schools To Open Or Close'' and ``Data Science To Diversify Learning and Reduce Commuting Times'' are significantly more likely to be opened than a control email subject line that reads ``Diversify Learning''---selected for its simplicity.  This is particularly interesting because these subject lines were informed by conversations with school board members in between campaign waves.  After observing low dashboard engagement following a few email campaign waves, we reached out to some of the board members who were included in the outreach to see if they might be willing to have a short conversation to share their views and feedback on our project.  We informally spoke with three board members across three different districts in two states.  The conversations were enlightening in many ways, but a few key takeaways emerged.  In one conversation, a school board member seemed surprised that it might be possible to redraw attendance boundaries in ways that might reduce segregation and travel times simultaneously, and believed others may find this to be surprising as well.  This motivated us to design the ``Data Science To Diversify Learning and Reduce Commuting Times'' subject line in an effort to create more of a ``hook'' to engage future recipients.  In another conversation, a school board member described the capacity planning challenges they faced as their district's population rapidly changed with immigration.  The board member described the need to determine which schools to open or close (and where) as the main impetus for their ongoing boundary planning efforts.  Even though their board cared about school diversity, on its own, the member highlighted it probably would not spark new boundary analyses.  The board member mentioned, however, that diversity might be an added consideration as a part of such ongoing boundary planning efforts, and that our tools and algorithms may help support such planning efforts by dovetailing issues of diversity with issues that are of more pressing concern for districts.  This prompted us to design the subject line ``Data Science To Decide Which Schools To Open Or Close'', in an attempt to foreground and lead with a concern that may be more top-of-mind (and less controversial) for school board members than issues of segregation and diversity.

The results in Figure~\ref{fig:2}(a) also show that the higher the percentage of elementary school student body in the district that is white, the more likely they are to open our emails (regardless of campaign), and that suburban district board members are slightly less likely to open our emails than those in urban districts---in both cases, after controlling for all other variables in the regression.

Figure~\ref{fig:2}(b) shows results for email click-throughs to the dashboard (N=4,320; results look virtually identical when conditioning on those who opened the email).  Despite the low number of click-throughs, the results suggest several associations between email campaigns, individual/ district/ campaign-level variables and click activity.  For one, after controlling for the same recipient and district-level characteristics as before, two campaigns offering \$20 to board members for their input on depicted boundaries generate \textit{lower} click-through rates than the control campaign with a subject line of ``Diversify Learning'' and no incentive.  While the body of the email specifies that the \$20 is meant to be donated to a school/teacher of their choosing, it is possible that board members are weary of clicking through for fear of being perceived as accepting a ``gift''---a politically precarious practice---for their participation.  Mentioning ``attendance boundaries'' in the subject line and body of the email also tends to generate lower click-through rates---perhaps because of how controversial and unpopular the topic is\footnote{As we sought his feedback on the dashboard, one former superintendent of a large suburban school district shared that changing attendance boundaries is so unpopular that it often costs school board members their jobs / opportunities for re-election}.  On the other hand, the emails that produce higher open rates (as described earlier) also appear to produce higher click-through rates.   

Importantly, we highlight these results as exploratory and directional, and far from conclusive.  Given the number of variables in our regressions, there is the possibility of an inflated Type I error rate (i.e., the likelihood of detecting significant associations purely by chance) due to multiple comparisons.  Indeed, conservatively applying Bonferonni correction to correct for the number of inferred coefficients in the email opens model renders all predictors insignificant at $\alpha = 0.05$.  Predictors remain significant in the email clicks model even after correction, but noting the small number of overall clicks, we still encourage readers to interpret these results as preliminary.  With these caveats in mind, we observe district leaders' engagement to be sensitive to how our outreach is framed and contextualized: district leaders appear more likely to open and click through on emails that use less controversial language that also appears to align with their a priori student assignment and planning objectives.  While these findings are interesting and may inform future outreach efforts for this and other projects, we note that the limited click-through rates coupled with even lower rates of dashboard exploration due to large dropoff at the IRB study description text render these cold email campaigns as ineffective channels for engaging school districts in considering educational data science-informed policies for fostering more diverse schools.

\subsection{Media coverage}
Observing low click-through rates and engagement with the dashboard, we turned to another channel for reaching school districts: media coverage.  There is a vast literature on media effects, including how media coverage can shape public discourse and opinion across a range of topics~\cite{king2017news}.  We reached out to an education journalist at the Hechinger Report, an outlet covering Educational Inequality and Innovation, for advice on how to gain more attention and engagement from district leaders.  Our outreach was ``almost cold'': we had only previously spoken with the journalist once, almost 18 months earlier, about a separate research project.  After demoing our dashboard to the journalist, they decided to write a story describing the tool and how the boundary scenarios were produced~\cite{barshay2022boundaries}.  Stumbling across this piece one week later, another journalist---a host of Georgia NPR's \textit{All Things Considered} show---also published a piece highlighting the tool and how districts might use it to inform their own diversity planning efforts~\cite{biello2022gpb}.  

We analyzed engagement with the dashboard in the 5 weeks following the first story's release to better understand how levels of exploration differ across different types of districts represented on the dashboard.  During this time, the dashboard received over 3,700 visits, which contributed to the exploration of boundary results for over 500 districts across 42 states.  Given extremely low levels of engagement with the dashboard prior to these stories, we believe it is safe to attribute increased engagement with the dashboard during this time period to the traffic that the stories likely drove.

\begin{figure}
\centering
\includegraphics[width=.9\linewidth]{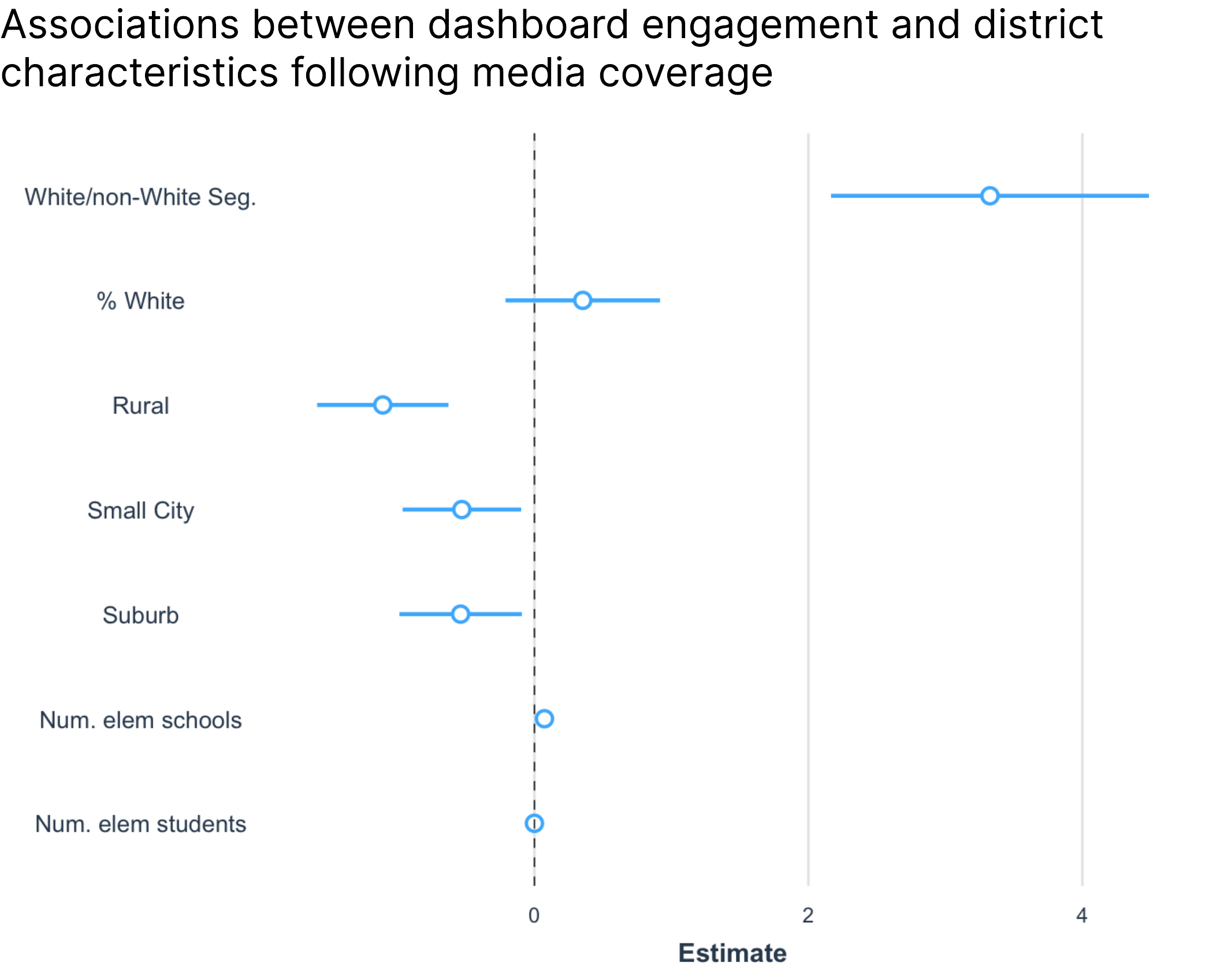}
\caption{Outputs of a negative binomial regression depicting associations between the number of times a particular district is explored on the dashboard and various district-level characteristics.}
\label{fig:3}
\end{figure} 

Figure~\ref{fig:3} shows the associations inferred from a negative binomial regression modeling the number of explorations per district on the dashboard as a function of various district-level characteristics.  Interestingly, urban districts were more likely to be explored than suburban, small city, or rural ones.  Furthermore, larger districts (defined in terms of the number of elementary schools in our data for them) were also more likely to be explored---likely because they have larger populations overall.  Perhaps most notably, the strongest predictor of whether or not a district would be explored, after controlling for all others, was its level of White/non-White segregation: more segregated districts were more likely to be explored.  This suggests that media coverage was effective in drawing attention to those districts where there is, arguably, great scope for change.

Given the broad nature of media coverage, it is likely that both district leaders and community members visited and engaged with the dashboard during the period following the articles' publishing.  Furthermore, given the anonymous nature of the media-driven dashboard visits, it is impossible to know who explored the dashboard in which ways, or from which districts.  Nevertheless, as the next section describes, a small set of people from both sets of stakeholders took time to explore and share their reflections and feedback on the premise and outputs of the boundary simulations. 

\subsection{Survey responses}
In the months following the email campaigns and media coverage (from May through October 2022), we collected approximately 11 responses via the feedback survey linked off of the dashboard.  Two survey responses came from email recipients; the remainder, we believe, came from both district leaders and community members who learned about the dashboard via the published articles.  Given that thousands of individuals explored the dashboard, 11 survey responses constitutes an extremely small---and hence, likely very biased---sample of total possible respondents.  Noting this selection bias, we caution against making generalized claims from the input.  Nevertheless, like interviews, we believe the qualitative insights provided by these survey responses shed invaluable light onto Research Question 2---``How do school district leaders and community members perceive these boundary changes---namely, what potential merits and pitfalls do they see in both how these boundaries were produced and their potential viability in practice?''.

When asked how happy they would be with the depicted rezonings, approximately an equal share of respondents expressed they would be somewhat/very happy and somewhat/very unhappy (three each), with two responding ``neutral''.  Interestingly, when asked how happy they believed other families would be with the depicted rezonings, only one respondent indicated ``somewhat happy''; eight indicated believing families would be somewhat or very unhappy.  This reflects potential selection biases in who filled out the survey (i.e., those who may be more open to systemic, equity-promoting changes in schools) and underscores the controversial nature of boundary changes across many school districts.  It may also point to potential clashes between student assignment models and the values of certain subsets of families~\cite{robertson2021vsd}.  When asked how much they trusted the computer program that produced the rezonings, five respondents indicated they sort of trusted it or trusted it completely, while three indicated they don't trust it, and two were unsure.  While these responses are mixed, it is interesting---and perhaps concerning---to see such high levels of trust, especially since the dashboard provided no insights into how the algorithm was working or which computational methods were applied to produce the rezonings (though the media coverage offered a high-level description).  Only one out of 9 respondents disagreed that changing attendance boundaries is a valuable strategy for promoting more diverse schools.  Finally, when asked how they felt about the idea of redrawing attendance boundaries to increase diversity in their district's schools after learning about this project, only one respondent said they were less interested in the idea than before.  An approximately equal number indicated the same or more interest as before (five and four, respectively), while one expressing same interest wrote in that they felt more ``...hopeful and inspired''.

Respondents' open-ended feedback offered a number of rich insights.  When asked what they found surprising, interesting, or novel about the depicted rezonings, several respondents shared interest in the possibility of fostering more diverse schools without significant travel disruptions.  For example, one respondent commented that they felt ``...Surprise that this could be done. Surprise that the transportation impact is so small.''  Similarly, another respondent---a school district official--said that they ``Love that it didn’t move Children to Far from their old School.''  Another school board member shared that ``I am interested to see that...there are improvements that could be made without major changes in time/distance to schools''.  One school board member shared an optimistic view, commenting that ``This is helpful information and a worthy project, and I don’t want perfect to be the enemy of improvement. I think there are practical solutions to improving racial diversity here.''

Yet respondents also had consistent suggestions for improvement.  When asked what they found troubling or concerning about the depicted rezonings, many commented on the fact that the depicted changes to school demographics were quite small.  One school board member mentioned ``The computer program only had 1 school out of 13 with a significant change which is not worth the headaches rezoning causes.  The other 12 barely changed if at all.''  Another respondent similarly questioned ``Is it worth disrupting families for a small change.'' Several other respondents highlighted the dashboard's failure to take into account the nuances and details of their unique local context when depicting hypothetical rezonings.  One respondent described the complexity of redrawing boundaries in the context of schools with special programs, sharing the following: ``I think the idea is great, but in a city like [redacted] there are other complexities. We have several schools that are language immersion, serving the local community. They are a 50/50 DLI [Dual Language Instruction] model. This really influences how boundaries can move and what the actual neighborhood space looks like. Additionally we have both k-5 and k-8 and this changes how boundaries can move.''  Another, a board member, commented on the algorithm's unrealistic school capacity constraints, highlighting that many of the schools that the algorithm allocated more students to were already at or over capacity (which is unknowable from the Department of Education's (DOE) Common Core of Data).  Some described other shortcomings, like missing schools (e.g. closed enrollment or newly built ones), a failure to account for measures of community cohesion (like elementary-middle-high feeder patterns), and other inaccuracies due to aggregated and dated DOE data.

Finally, we found it encouraging that several respondents shared their contact information to continue the conversation and share additional feedback.  One board member wrote that they ``...would like to chat with [us] about updating [our] maps to include more accurate information about the schools'', ending their note with the following question: ``Are you able to talk with me sometime???''  One district official expressed interest in seeing results for other schools, asking ``When will the middle and high school data update''?  Finally, a community member wrote asking ``How can I work to show this to my county board and persuade them?''  Despite the myriad of possible improvements and iterations that would be required to make the depicted rezonings applicable in practice, there appeared to be interest from at least a handful of leaders and members across districts.  We have followed up with respondents to learn more about their unique contexts and are currently collaborating with districts to translate the tools and methods from this research into practice.

\section{Discussion}

Recall our original research questions:

\begin{enumerate}
    \item Which messaging and outreach strategies are more or less effective in prompting school district leaders and community members in exploring potential boundary changes that could help foster more racially/ethnically integrated schools?
    \item How do school district leaders and community members perceive these boundary changes---namely, what potential merits and pitfalls do they see in both how these boundaries were produced and their potential viability in practice?
\end{enumerate}

With respect to question 1, while we find that email campaigns using subject lines that align diversity goals with other perennial district-level concerns (like school capacity planning or travel times) yield larger open and click-through rates, cold email outreach proves largely ineffective as a means for prompting district leaders to explore hypothetical diversity-promoting boundaries.  Media coverage, however, generates much more exploration---particularly among scenarios for districts that are more segregated.  We cannot say that media coverage is a more efficient method for sharing educational data science research findings with districts; indeed, the greater volume of exploration it prompts may simply be due to its wider reach.  Nevertheless, it is clear that in this particular setting, media coverage resulting from (almost cold) outreach to an education journalist produced a level of engagement that emails alone could not accomplish.  We acknowledge several limitations in our email campaign and analysis.  For one, given the low engagement volume (especially click-through rates) and differences in engagement rates for emails with similar subject lines, it is important to interpret all results as exploratory and suggestive.  Replication efforts may help illuminate the robustness of results, as might follow-on analyses that account for other variables that could have introduced noise or otherwise affected outcomes.  While click-through rates were low, the high open rates offer an interesting direction for future work.  For example, instead of asking education stakeholders to click through to a dashboard, the dashboard and/or its most salient findings may be brought into the body of the email---increasing the likelihood that recipients actually see it.

With respect to question 2, an important caveat of our largely survey response-driven findings to this question is the extremely small number of responses (11), especially when compared to the thousands of district data explorations that occurred through the dashboard.  We did not hear from the vast majority of those who explored the dashboard, and so, cannot make general claims about their perceptions of specific boundary changes.  Nevertheless, those who did respond generally indicated an interest in the idea of redrawing school attendance boundaries to foster more diverse schools, and several asked to have follow-on conversations to explore how these preliminary research findings might be applied in their local settings.  Many also had constructive suggestions for improving the algorithms to make their outputs more useful in practice.  Admittedly, even though the response pool was small, given how contentious and unpopular boundary changes are, we were surprised and encouraged that \textit{anyone} responded affirmatively at all.  Our findings thus suggest that there is hope for making progress on a problem as entrenched as racial and ethnic segregation in schools, through means as controversial as attendance boundary changes---even if just in a few school districts to begin with.  Indeed, sparking change in a few districts serving thousands or tens of thousands of students can translate into scalable impacts on learning and future life outcomes for students.

Looking ahead, we believe there are two under-explored opportunities for educational data scientists to consider pursuing as they shape their research agendas in the months and years ahead.  The first involves conducting research at the intersection of computation and \textit{education policy}---particularly focusing on systemic issues like segregation---in addition to the myriad of efforts underway exploring the intersection of computation and \textit{learning}.  The growth of educational data science practices like ``learning engineering''~\cite{thille2016learning} offer new and exciting opportunities for using computational methods to advance academic mastery, socioemotional understanding, and various other cognitive and metacognitive skills.  Yet learning is often shaped by political and sociological forces before students even step foot in a school or classroom.  Through computational research projects like boundary modeling and other policy-related efforts, educational data scientists can help study and perhaps even shape these ``upstream'', systemic forces that can silently, yet powerfully, affect the opportunities afforded to students and families.  The second opportunity involves exploring which communication and engagement strategies are most effective for translating research findings into practice.  Too often, the fruits of great research remain unrealized because the paths to successful implementation are unclear.  We believe there is a rich opportunity for educational data scientists to explore intersections between their work and ``implementation science'' in order to ensure that promising early research is translated and further evaluated across different contexts.  We hope our study offers a useful, albeit preliminary, reference point for both of these exciting future directions.

\section{Acknowledgements}
Funding for this project was provided by the MIT Center for Constructive
Communication. The authors declare no competing interests. The research was approved by MIT’s Institutional Review Board.  We are grateful to Peter Bergman, Kumar Chandra, Alvin Chang, Akeshia Craven-Howell, Rebecca Eynon, Tad Hirsch, Melissa Krull, Justin Reich, Todd Rogers, Joshua Starr, district leaders who responded to our outreach, and the anonymous reviewers for their helpful comments and guidance on this project.  We also thank Jill Barshay and Peter Biello for their support in sharing the research through press channels.  Finally, we thank Lauren J. Bierbaum and Max Vargas at the XQ Institute for sharing their input and enabling us to access public records for elected school board members.  

% \begin{acks}

% Alvin, Rebecca, Justin, XQ institute for data, Kumar, Todd Rogers
% IRB approval
% \end{acks}

\bibliographystyle{ACM-Reference-Format}
\bibliography{sample-base}

\appendix

\section{Email subject lines and corresponding hypotheses}
\label{app:subjects}
Table~\ref{tab:app_subject} shows the different subject lines, motivating hypotheses, and waves these emails were sent during.

\begin{table}[ht]
    \centering
    \begin{tabular}{p{0.35\linewidth} | p{0.5\linewidth} | p{0.15\linewidth}}
      \textbf{Email subject line} & \textbf{Hypothesis} & \textbf{Wave(s)} \\ \hline
      Diversify Learning & Concise subject line that highlights project topic & Test, 1\\
      Diversify Learning and get \$20 for a school & A small incentive that board members can allocate to a teacher of their choice may increase engagement & 1\\
      Data science to diversify attendance boundaries & ``Data science'' might pique interest, especially in the context of boundary planning & 2\\
      \$20 for your views on diversifying attendance boundaries & A small incentive that board members can allocate to a teacher of their choice may increase engagement & 2\\
      Data science to improve learning experiences & ``Data science'', focusing on learning instead of diversity may increase engagement & 2 \\
      Data science to diversify learning experiences & Same as above, but mentioning diversity and learning together & 2, 5 \\
      More diversity with shorter commutes? & Stating the main finding of~\cite{gillani2023redrawing} up front may increase engagement & 3, 5 \\
      More diversity with shorter commutes? \$20 for your thoughts! & Stating the main finding of~\cite{gillani2023redrawing} up front and adding the incentive may increase engagement & 3 \\
      More diversity, less driving? & Pithy description of findings from~\cite{gillani2023redrawing} & 4 \\
      2,000 pennies for your thoughts & Pithy version of incentive email & 4 \\
      Improve discussions about school diversity with families & Heard from districts that they find it difficult to talk with families about diversifying schools & 5 \\
      Data science to diversify learning and reduce commuting times & Combing learning, diversity, and shorter commute messages from earlier & 6 \\
      Data science to decide which schools to open or close & Framing in terms of a problem school districts often face that triggers boundary planning may increase engagement & 6 \\
      Data science to diversify learning and reduce commute times & Has elements of promising subject lines throughout runs & Final \\
    \end{tabular}
    \caption{Email subject lines and corresponding hypotheses motivating their inclusion across different email waves.}
    \label{tab:app_subject}
\end{table}

\section{Dashboard survey questions}
\label{app:survey}

The optional survey linked off of the dashboard included the following questions:

\begin{itemize}
    \item How happy would you be with this rezoning? [Very happy / Somewhat happy / Neutral / Somewhat unhappy / Very unhappy / Other:]
    \item How happy do you believe other families would be with this rezoning? [Very happy / Somewhat happy / Neutral / Somewhat unhappy / Very unhappy / Other:]
    \item How much do you trust the computer program that created this rezoning? [I trust it completely / I sort of trust it / I'm not sure if I trust it or not / I don't trust it / Other:]
    \item After learning about this project, how do you feel about the idea of redrawing attendance boundaries to increase diversity in your district's schools? [More interested in the idea than before / Less interested in the idea than before / About the same level of interest as before / Other:]
    \item How much do you agree with the following statement: "I believe changing attendance boundaries is a valuable strategy for promoting more diverse schools in my district" [Strongly agree / Agree / Neutral / Disagree / Strongly disagree / Other:]
    \item What's something you found interesting, surprising, or novel about the proposed rezoning, if anything? [Open-ended response]
    \item What's something you found troubling or concerning about the proposed rezoning, if anything? [Open-ended response]
    \item Do you have any other thoughts or reactions that you'd like to share? [Open-ended response]
    \item Please share your contact information if you would like for a member of our team to reach out to you to discuss this project further, including how it might be applied to your school district.
\end{itemize}

\section{Dashboard landing page IRB text}
\label{app:irb}

Email recipients who clicked through saw the following text on the dashboard landing page.\\

\begin{quote}
\textit{Hello!  We are scientists at the Massachusetts Institute of Technology, and we've found that it may be possible to increase diversity across many districts’ schools without creating large disruptions for families.  We invite school district leaders like yourself to explore these results and share your feedback through this website, which is part of a research study we are conducting in order to better understand how education leaders respond to different diversity policies.  Exploring the dashboard is completely voluntary, and you can stop anytime.  We don’t expect any discomforts from doing so, and hope you’ll find our results interesting and informative for your work.  Additionally, there will be no public record or knowledge of your participation, so you can feel free to engage (or not engage!) in a way that’s best for you.  If you have any questions or comments, feel free to email us at schooldiversity@media.mit.edu or contact MIT's research office at couhes@mit.edu.}
\textit{Please click “Continue” to get started!}
\end{quote}

\end{document}